\documentclass[12pt,english]{iopart}

\usepackage{iopams}
\usepackage{setstack}
\usepackage{amstext}
\usepackage{babel}
\usepackage{graphicx}

\begin{document}

\global\long\def\pd{\partial}
\global\long\def\kiss#1#2{\left|#1\right\rangle \left\langle #2\right|}
\global\long\def\ket#1{\left|#1\right\rangle }
\global\long\def\bra#1{\left\langle #1\right|}
\global\long\def\avg#1{\left\langle #1\right\rangle }
\global\long\def\abs#1{\left|#1\right|}
\global\long\def\Tr{\textrm{Tr}}
\global\long\def\Id{\textrm{Id}}
\global\long\def\a{\alpha}
\global\long\def\b{\beta}
\global\long\def\m{\mu}
\global\long\def\n{\nu}
\global\long\def\t{\theta}
\global\long\def\d{\delta}
\global\long\def\e{\epsilon}
\global\long\def\g{\gamma}
\global\long\def\r{\rho}
\global\long\def\s{\sigma}
\global\long\def\l{\lambda}
\global\long\def\o{\omega}
\global\long\def\k{\kappa}
\global\long\def\z{\zeta}
\global\long\def\P{\Psi}
\global\long\def\G{\Gamma}
\global\long\def\D{\Delta}
\global\long\def\L{\Lambda}
\global\long\def\O{\Omega}
\global\long\def\S{\Sigma}
\global\long\def\dd{\dagger}
\global\long\def\ua{\uparrow}
\global\long\def\da{\downarrow}
\global\long\def\ra{\rightarrow}
\global\long\def\la{\leftarrow}
\global\long\def\x{\times}
\global\long\def\kp{\otimes}

\title{On Coulomb drag in double layer systems}

\author{Bruno Amorim$^1$ and N M R Peres$^2$}
\address{$^1$ Instituto de Ciencia de Materiales de Madrid,
CSIC, Cantoblanco, E-28049 Madrid, Spain.}
\address{$^2$ Physics Department and CFUM, University of Minho, P-4710-057, Braga, Portugal.}
\ead{amorim.bac@icmm.csic.es}

\date{\today}

\begin{abstract}
We argue, for a wide class of systems including graphene, that in the low temperature,
high density, large separation and strong screening limits the drag
resistivity behaves as $d^{-4}$, where $d$ is the separation between
the two layers. The results are independent of the energy dispersion relation, the
dependence on momentum of the transport time, and the electronic wave function structure.
We discuss how a correct treatment of the
electron-electron interactions in an inhomogeneous dielectric background
changes the theoretical analysis of the experimental drag results 
of Kim \textit{et al} (2011 \textit{Phys. Rev.} \textbf{B} 161401). We find that a quantitative understanding of the 
available experimental data  (Kim \textit{et al} 2011 \textit{Phys. Rev.} \textbf{B} 161401) for drag in graphene is lacking.
\end{abstract}


\maketitle

\section{Introduction}

Coulomb drag \cite{Pogrebinskii1977,Price1983} occurs when driving a 
current through a metallic layer, referred to as the active layer 
(and which will be denoted by $2$), induces a current in another metallic 
layer, separated by a distance $d$, referred to as the passive layer 
(and which  will be denoted by $1$). This phenomenon is caused by the 
transference of momentum between electrons in different layers
 due to the interlayer electron-electron interaction. 
In experimental situations no current is allowed to flow
 in the passive layer, so that an electrical field, $E_{1}$,
 builds up in that layer. In this situation, the 
drag resistivity, $\rho_{D}$, is obtained from the ratio
\begin{equation}
\rho_{D}=\frac{E_{1}}{j_{2}},
\end{equation}
where $j_{2}$ is the current driven through the active layer.

Recently there has been great deal of interest on the phenomenon of 
Coulomb drag in graphene double layers. Although the number of experimental 
works on drag in graphene is still scarce \cite{Kim2011,Geim2012,Kim2012}, 
there are already plenty of theoretical works on the 
topic \cite{Narozhny2007,Tse2007,Peres2011,Katsnelson2011,Hwang2011,Narozhny2012,Polini2012}.
However, among the theoretical literature we find contradictory
 statements, particularly in the limit of low temperature, 
high density, large layer separation and strong screening. 
In this limit, it is a well established result that the drag 
resistivity between two 2DEG's (two dimensional electron gases), with parabolic dispersion 
relations and a constant transport time, depends on 
temperature as $T^{2}$ and on the interlayer 
separation as $d^{-4}$ \cite{Smith1993} 
(a result previously given in \cite{Gramilla1991} 
without derivation). Graphene is distinct from the 2DEG in 
three ways: (i) the low energy dispersion relation is linear, 
instead of parabolic; (ii) the electron wave function has a 
spinorial structure, instead of being a scalar; 
(iii) the transport time for the dominant kind of 
impurities is proportional to the momentum, $\tau(k) \propto k$, \cite{Peres2010Review}.
 There is agreement \cite{Tse2007,Peres2011,Katsnelson2011,Hwang2011,Narozhny2012,Polini2012} 
that in the low temperature/high density limit, the $T^2$ 
dependence should still hold in graphene. However, there are a few contradictory
results on the dependence of the drag resistivity on the layer 
separation, and how considering a constant or a momentum dependent 
transport time changes the result. In this paper we will give special 
attention to the limit of low temperature, high density, large interlayer 
separation and strong screening. Therefore, we will make a brief overview of 
the available results in the literature in this limit.

\begin{enumerate}
\item Tse {\it et al} \cite{Tse2007} assumes a constant transport time, and a dependence of $d^{-4}$ is obtained.

\item Peres {\it et al} \cite{Peres2011} considers a momentum dependent transport time, $\tau(k) \propto k$, obtaining a $d^{-6}$ dependence
(as we explain later this result is due to an algebraic error at the end of the asymptotic calculation; correcting this gives a $d^{-4}$ dependence).

\item Katsnelson \cite{Katsnelson2011} considers a constant transport time and obtains a $d^{-4}$ dependence.

\item Hwuang {\it et al} \cite{Hwang2011} considers both cases of a constant transport time and a transport time proportional to the momentum, $\tau(k) \propto k$. For a constant transport time a $d^{-6}$ dependence is obtained (in contradiction with the result from \cite{Tse2007}). For the case of a momentum dependent transport time a $d^{-4}$ behaviour is obtained. The case of drag between two bilayer graphene layers is also studied. Using a constant transport time a $d^{-4}$ behaviour is obtained in this same limit.

\item Narozhny {\it et al} \cite{Narozhny2012} considers both cases of a constant and a linearly momentum dependent transport time. For the case of a constant transport time it is obtained a $d^{-4}$ dependence. It is argued that in the low temperature, high density limit, this result still holds, regardless of the momentum dependence of the transport time.

\item Carrega {\it et al} \cite{Polini2012}, in a recent independent work, studies drag between massless Dirac electrons. It is proved that in the low temperature/high density limit, the dependence on momentum of the transport time is irrelevant and a $d^{-4}$ dependence is obtained for large interlayer separations.
\end{enumerate}

In this paper, we attempt to clarify this situation by presenting a clear proof that in the limit of low temperature, high density, large interlayer separation and strong screening the drag resistivity should always depend on temperature as $T^{2}$ and on distance separation as $d^{-4}$. Our analysis is independent of the energy dispersion relation, electron wave function structure and dependence on momentum of the transport time. The structure of the paper is as follows: in \sref{sec:Formulation}, we present the general theory of Coulomb drag. In \sref{sec:Universal}, we present a general argument proving that the drag resistivity in the limit of low temperature, high density, large separation between layers and strong screening should always depend on temperature as $T^{2}$ and on the interlayer separation as $d^{-4}$. We also study the case of graphene for small interlayer separation. In \sref{sec:Drag-in-graphene}, we specialize to graphene and re-analyze the asymptotic result derived in \cite{Peres2011}. Finally, in \sref{sec:Experiments} we look into the experimental data of \cite{Kim2011} using a more careful treatment of the bare Coulomb interactions.

\section{General formulation of Coulomb drag}\label{sec:Formulation}

Theoretically, it is  convenient to compute conductivities instead of resistivities. For isotropic systems
the drag resistivity is related to the conductivities by
\begin{equation}
\rho_{D}=-\frac{\sigma_{12}}{\sigma_{11}\sigma_{22}-
\sigma_{12}\sigma_{21}}\simeq-\frac{\s_{12}}{\sigma_{11}\sigma_{22}},\label{eq:rho_sigma}
\end{equation}
where $\sigma_{11}$ and $\sigma_{22}$ are the intralayer
conductivities of the passive and active layers and $\sigma_{D}\equiv\s_{12}=\sigma_{21}$
is the drag conductivity; in \eref{eq:rho_sigma} it was
assumed that $\sigma_{11},\sigma_{22}\gg\sigma_{D}$.

Considering that  tunnelling between layers does not occur  and that the
intralayer transport is dominated by impurity scattering, the drag conductivity
can be computed in second order in the interlayer interaction using either Boltzmann's kinetic equation \cite{Hwang2011,Peres2011,Smith1993,Flensberg1995a}, the memory-function formalism \cite{Zheng1993}
or  Kubo's formula \cite{Hwang2011,Narozhny2012,Kamenev1995b}, and
is given by
\begin{equation}
\sigma_{D}^{ij}=\frac{e_{1}e_{2}}{16\pi k_{B}T}\int\frac{d^{2}q}{\left(2\pi\right)^{2}}\int_{-\infty}^{+\infty}\frac{d\o \abs{U_{12}(\vec{q},\o)}^{2}}{\sinh^{2}\left(\beta\hbar\omega/2\right)}\Gamma_{1}^{i}(\vec{q},\o)\Gamma_{2}^{j}(\vec{q},\o),\label{eq:SigmaDrag}
\end{equation}
where $e_{a}$ is the charge of carriers in layer $a$, $U_{12}(\vec{q},\o)$
is the interlayer electron-electron interaction and $\Gamma_{a}^{i}(\vec{q},\o)$
is the $i$-th component of non-linear susceptibility of layer $a$.
In the weak impurity limit, the non-linear susceptibility reads \cite{Tse2007,Hwang2011} (the
layer index is omitted for simplicity)
\begin{eqnarray}
\fl \vec{\G}(\vec{q},\o)=-2\pi g\sum_{\l,\l^{\prime}}\int\frac{d^{2}k}{\left(2\pi\right)^{2}}
f_{\vec{k},\vec{k}+\vec{q}}^{\l,\l^{\prime}}\left[n_{F}(\e_{\vec{k},\l})-n_{F}(\e_{\vec{k}+\vec{q},\l^{\prime}})\right]\nonumber\\
\times \left(\vec{v}{}_{\vec{k},\l}\tau_{\vec{k},\l}-\vec{v}{}_{\vec{k}+\vec{q},\l^{\prime}}\tau_{\vec{k}+\vec{q},\l^{\prime}}\right)\d\left(\e_{\vec{k},\l}-\e_{\vec{k}+\vec{q},\l^{\prime}}+\hbar\o\right),\label{eq:NonLinPol}
\end{eqnarray}
where $g$ is the flavour degeneracy, $\l,\lambda^{\prime}$ are band
indices, $f_{\vec{k},\vec{k}^{\prime}}^{\l,\l^{\prime}}=\left|\left\langle \vec{k},\l\mid\vec{k}^{\prime},\l^{\prime}\right\rangle \right|^{2}$
is the electron wave function overlap factor (which encodes the structure of the wave function), $n_{F}(\e)$ is the Fermi-Dirac
distribution function, $\vec{v}{}_{\vec{k},\l}$ is the particle's group
velocity, $\tau_{\vec{k},\l}$ is the impurity transport time, and
$\e_{\vec{k},\l}$ is the energy dispersion. 

For the interlayer interaction one usually uses the RPA dynamically
screened Coulomb interaction \cite{Flensberg1995a},
\begin{equation}
U_{12}(\vec{q},\omega)=\frac{V_{12}(\vec{q})}{\epsilon_{RPA}(\vec{q},\omega)},
\end{equation}
where $V_{ab}(\vec{q})$ is the bare Coulomb interaction between electrons
in layer $a$ and $b$, and $\epsilon_{RPA}(\vec{q},\o)$ is the RPA
dielectric function for the double layer system, which is given by \cite{Flensberg1995a,DasSarmaApendixA1993}
\begin{eqnarray}
\fl \e_{RPA}(q,\omega)=\left[1-V_{11}(\vec{q})\chi_{1}(\vec{q},\o)\right]\left[1-V_{22}(\vec{q})\chi_{2}(\vec{q},\o)\right]\nonumber\\
-V_{12}(\vec{q})V_{21}(\vec{q})\chi_{1}(\vec{q},\o)\chi_{2}(\vec{q},\o),
\end{eqnarray}
and $\chi_{a}(\vec{q},\o)$ is the polarizability of layer $a$. The
bare Coulomb interactions can in general be written as (see \ref{sec:Coulomb_Interaction})
\begin{equation}
V_{ab}(\vec{q}) = \frac{1}{\e_{ab}(q)}\frac{e^{2}}{2\e_{0}q}e^{-qd\left(1-\delta_{ab}\right)},\label{eq:bareCoulomb}
\end{equation}
where $\e_{0}$ is the vacuum permittivity and $\e_{ab}(q)$ are effective
dielectric functions. If the metallic layers are immersed in a homogeneous dielectric with constant $\e_{r}$ then $\e_{ab}(q)=\e_{r}$.

\section{Low temperature behaviour\label{sec:Universal}}

We  now study the behaviour of the drag conductivity in the limit
of low temperature, $\e_{F1(2)}\gg k_{B}T$, where $\e_{F1(2)}$ is
the Fermi energy of layer $1(2)$. Unless specified otherwise, we
will keep the energy dispersion relation, $\e_{\vec{k},\l}$, the
transport time, $\tau_{\vec{k},\l}$, and wave function overlap factors,
$f_{\vec{k},\vec{k}+\vec{q}}^{\l,\l^{\prime}}$, general. We assume 
isotropy and that there is only one band at the
Fermi level. 
Central to the analysis it the realization that
the energy dispersion relation close to the Fermi energy is always linear
in momentum, that is, the dispersion can be approximated by:
\begin{equation}
\e_{\vec{k},c}-\e_{F}\simeq\hbar v_{F}\left(k-k_{F}\right)\,,
\label{eq:dispersion}
\end{equation}
where $v_{F}$ is the slope of the band at the Fermi energy,
termed Fermi velocity, 
and the label $c$ refers to the conduction band. 
For graphene \eref{eq:dispersion} is exact.
We  also assume that the two metallic layers are placed in vacuum,
such that
\begin{eqnarray}
\label{eq:RPA_vac}
\fl \e_{RPA}(q,\o)=1+\chi_{1}(\vec{q},\o)\chi_{2}(\vec{q},\o)\left(\frac{e^{2}}{2\e_{0}q}\right)^{2}2\sinh\left(qd\right)e^{-qd}\nonumber\\
-\frac{e^{2}}{2\e_{0}q}\left[\chi_{1}(\vec{q},\o)+\chi_{2}(\vec{q},\o)\right].
\end{eqnarray}

Due to the factor $\sinh^{-2}\left(\beta\hbar\omega/2\right)$ in \eref{eq:SigmaDrag},
the main contribution to the integral in $\o$ comes from $\hbar\o\lesssim k_{B}T$.
Since $\e_{F1(2)}\gg k_{B}T$, we can therefore expand the remaining
integration kernel to lowest order in $\o$ and set $T=0$. Therefore
we replace the dynamically screened dielectric function, $\e_{RPA}(\vec{q},\o)$,
by its static value, $\e_{RPA}(\vec{q},0)$, and expand the non-linear
susceptibility of each layer, \eref{eq:NonLinPol}, to lowest order
in $\o$. Using the energy conserving $\d$-function in \eref{eq:NonLinPol}, $\d\left(\e_{\vec{k},\l}-\e_{\vec{k}+\vec{q},\l^{\prime}}+\hbar\o\right)$, we expand to lowest order in $\o$:
\begin{eqnarray}
\fl n_{F}\left(\e_{\vec{k},\l}\right)-n_{F}\left(\e_{\vec{k}+\vec{q},\l^{\prime}}\right)=n_{F}\left(\e_{\vec{k},\l}\right)-n_{F}\left(\e_{\vec{k},\l}+\hbar\o\right)\nonumber\\
\simeq-\hbar\o\frac{\pd n_{F}\left(\e_{\vec{k},\l}\right)}{\pd\e}\simeq\hbar\o\d\left(\e_{F}-\e_{\vec{k},\l}\right).
\end{eqnarray}
Therefore $\vec{\G}(\vec{q},\o)$ has a linear contribution in $\o$.
Since we want $\vec{\G}(\vec{q},\o)$ to  lowest order in $\o$
we can now set $\o=0$ in $\d\left(\e_{\vec{k},\l}-\e_{\vec{k}+\vec{q},\l^{\prime}}+\hbar\o\right)$,
obtaining
\begin{eqnarray}
\fl \vec{\G}(\vec{q},\o)=-g\frac{\hbar\o}{2\pi}\int d^{2}kf_{\vec{k},\vec{k}+\vec{q}}^{c,c}\d\left(\e_{F}-\e_{\vec{k},c}\right)\d\left(\e_{F}-\e_{\vec{k}+\vec{q},c}\right)\nonumber\\
\times \left(\vec{v}{}_{\vec{k},c}\tau_{\vec{k},c}-\vec{v}{}_{\vec{k}+\vec{q},c}\tau_{\vec{k}+\vec{q},c}\right).
\end{eqnarray}
Since we have isotropy we can write $\vec{v}{}_{\vec{k},c}\tau_{\vec{k},c}=\vec{k}g(k)$,
where $g(k)$ is a general function that satisfies $k_{F}g(k_{F})=v_{F}\tau_{F}$,
 $\tau_{F}$ being the transport time at the Fermi level. The $\d$-functions
set $\left|\vec{k}\right|=\left|\vec{k}+\vec{q}\right|=k_{F}$, and
therefore we can take $g(k)$ outside the integral, obtaining to lowest
order in $\o$
\begin{equation}
\vec{\G}(\vec{q},\o)=g\frac{\hbar\o v_{F}\tau_{F}}{2\pi k_{F}}\vec{q}\int d^{2}k\, f_{\vec{k},\vec{k}+\vec{q}}^{c,c}\d\left(\e_{F}-\e_{\vec{k},c}\right)\d\left(\e_{F}-\e_{\vec{k}+\vec{q},c}\right).
\end{equation}
Note that in this limit $q$ is restricted to $q<2k_{F}$. To perform
the integration in $\vec{k}$, we choose, without loss of generality,
$\vec{q}=(q,0)$  and write 

\begin{eqnarray*}
u \equiv \cos\theta=\frac{\vec{k}\cdot\vec{q}}{kq},\\
\int d^{2}k = 2k_{F}\int_{0}^{\infty}dk\int_{-1}^{1}\frac{du}{\sqrt{1-u^{2}}},\\
\d\left(\e_{F}-\e_{\vec{k},c}\right) = \frac{1}{\hbar v_{F}}\d\left(k-k_{F}\right),\\
\d\left(\e_{F}-\e_{\vec{k}+\vec{q},c}\right) = \frac{1}{\hbar v_{F}q}\d\left(u+\frac{q}{2k_{F}}\right).
\end{eqnarray*}
Therefore the following result is obtained
\begin{equation}
\vec{\G}(\vec{q},\o)=g\frac{\o\tau_{F}}{\pi\hbar v_{F}}\frac{\vec{q}}{q}\left[\frac{f_{\vec{k},\vec{k}+\vec{q}}^{c,c}}{\sqrt{1-u^{2}}}\right]_{k=k_{F},u=-\frac{q}{2k_{F}}}.\label{eq:NLP_small_omega}
\end{equation}
This result is central to this paper. It shows that in the limit of low temperature and high density, the non-linear susceptibility is independent of both the energy dispersion relation and the dependence of the transport time on momentum, depending only in the particular form of the overlap factor. Therefore, this result can be readily applied for the case of a 2DEG, graphene, bilayer graphene and other systems. Although it was already pointed out in \cite{Narozhny2012,Polini2012} that the non-linear susceptibility is independent of the momentum dependence of the transport time, in those works this result was obtained for the particular case of massless Dirac electrons. Here, we show that this is a general result also independent of the energy dispersion relation. Note that this result contradicts \cite{Hwang2011}, where different
results for the non-linear susceptibility are obtained for different transport
times in the low temperature limit. 

Since in the low temperature limit
we have $\vec{\G}(\vec{q},\o)\propto\o$, the integration  in
$\o$ in \eref{eq:SigmaDrag} reads
\begin{equation}
 \int_{0}^{\infty}\frac{d\o\o^{2}}{\sinh^{2}\left(\beta\hbar\omega/2\right)}=
2^{3}\left(\frac{k_{B}T}{\hbar}\right)^{3}\frac{\pi^{2}}{6}\,,
\end{equation}
which gives the $T^{2}$ dependence of the drag conductivity and resistivity
in the low temperature limit. The $T^{2}$ behaviour is independent
of the details of the energy dispersion relation, the transport time
and the wave function overlap factors. Notice, however, that the $T^{2}$
behaviour might be modified if one includes corrections to the drag
conductivity due to higher order terms in the interlayer interaction \cite{Levchenko2008}.

\subsection{General system at large interlayer distance and strong screening}

We now  assume that the interlayer separation is large, $k_{F}d\gg1$.
The interlayer Coulomb interaction decays exponentially with  $d$,
thus  the integration kernel of \eref{eq:SigmaDrag}
is dominated by values of $q$ such that $q\lesssim d^{-1}$. Therefore
the condition $k_{F}d\gg1$ allow us to expand the remaining integration
kernel to lowest order in $q$. To lowest order,  the overlap
factor $f_{\vec{k},\vec{k}+\vec{q}}^{c,c}$ is $1$. Therefore at
low temperature and for small $q$ and $\o$, with $\o<v_{F}q$, we have
\begin{equation}
\vec{\G}(\vec{q},\o)=g\frac{\o\tau_{F}}{\pi\hbar v_{F}}\frac{\vec{q}}{q},\label{eq:NLPsmall_q_omega}
\end{equation}
a universal result that is independent of all the details of the system. Note, that although it is clear that in this limit $\vec{\Gamma}$ should only depend on quantities defined at the Fermi level ($k_{F}$, $\tau_{F}$), it is not obvious at first that changing the momentum dependence of $\e_{\vec{k}}$ or $\tau_{\vec{k}}$ will not change the momentum dependence of $\vec{\Gamma}$.  For small $q$ we approximate $\chi_{a}(q,0)\simeq-\rho_{a}\left(\e_{Fa}\right)$,
where $\r_{a}(\e)$ is the density of states of layer $a$, and the
RPA dielectric function \eref{eq:RPA_vac} becomes
\begin{equation}
\e_{RPA}(q,0)=1+\frac{q_{TF1}q_{TF2}}{q^{2}}2\sinh\left(qd\right)e^{-qd}+\frac{q_{TF1}+q_{TF2}}{q},\label{eq:Thomas-Fermi RPA}
\end{equation}
with $q_{TFa}=\rho_{a}(\e_{Fa})e^{2}/(2\e_{0})$, the Thomas-Fermi
screening momentum in 2D of layer $a$. If we assume that we have
strong screening, $q_{TF1(2)}d\gg1$, we further approximate \cite{Smith1993}
\begin{equation}
\e_{RPA}(q,0)=2\frac{q_{TF1}q_{TF2}}{q^{2}}\sinh\left(qd\right)e^{-qd}.\label{eq:RPA_StrongScreening}
\end{equation}
If the dispersion relation of layer $a$ is given by a power law,
$\e_{\vec{k},c}^{a}=C_{a}k^{\b_{a}}$, then we have $q_{TFa}\propto k_{Fa}^{2-\b_{a}}$.
Therefore, for a linear dispersion relation the condition $q_{TFa}d\gg1$ is  equivalent to $k_{Fa}d\gg1$; while for a parabolic dispersion
relation $q_{TFa}$ is independent of $k_{Fa}$, and
therefore $q_{TFa}d\gg1$ becomes an extra assumption. Assuming $q_{TF1(2)}d\gg1$,
and using \eref{eq:NLPsmall_q_omega} and \eref{eq:RPA_StrongScreening}
in \eref{eq:SigmaDrag} we obtain the following expression for the
drag conductivity: 

\begin{equation}
\s_{D}=\frac{e_{1}e_{2}}{\hbar}\frac{\zeta(3)g_{1}g_{2}}{2^{4}}\frac{e_{1}^{2}}{4\pi\e_{0}v_{F1}\hbar}\frac{e_{2}^{2}}{4\pi\e_{0}v_{F2}\hbar}\frac{\tau_{F1}\tau_{F2}\left(k_{B}T\right)^{2}}{\hbar^{2}\left(q_{TF1}d\right)^{2}\left(q_{TF2}d\right)^{2}}.
\label{eq:SigmaUniversal}
\end{equation}
This expression is valid for $\e_{F1(2)}\b,\, k_{F1(2)}d,\, q_{TF1(2)}d\gg1$
and is universal in the sense that is does not depend on the particular
forms of the energy dispersion relations, transport time dependence
on momentum or wave function structure. We obtain the familiar 2DEG $T^{2}$ and
$d^{-4}$ behaviour for the drag conductivity, proving that it is indeed a much more general result. If the metallic layers
are immersed in a homogeneous dielectric medium, with dielectric constant $\e_{r}$,
one should multiply \eref{eq:SigmaUniversal} by $\e_{r}^{2}$. Now, we notice that in the low temperature limit the intralayer conductivity for isotropic systems is given by the Boltzmann result 
\begin{equation}
\s_{aa}=\frac{e_{a}^{2}}{2}\r(\e_{Fa})v_{F}^{2}\tau_{F},
\end{equation}
where the factor of $1/2$ comes from the fact that we are in two dimensions,
the density of states at the Fermi energy is given by 
\begin{equation}
\r(\e_{Fa})=\frac{g_{a}}{2\pi}\frac{k_{Fa}}{\hbar v_{Fa}},
\end{equation}
and that the carrier density is related to the Fermi momentum in two
dimensions by
\begin{equation}
k_{Fa}=\sqrt{\frac{4\pi n_{a}}{g_{a}}}.
\end{equation}
This allow us to express the drag resistivity in terms of the carrier densities as
\begin{equation}
\rho_{D}=-\frac{\hbar}{e_{1}e_{2}}\frac{\zeta(3)}{2^{6}\pi\sqrt{g_{1}g_{2}}}\left(\frac{4\pi\e_{0}}{e_{1}^{2}}\right)\left(\frac{4\pi\e_{0}}{e_{2}^{2}}\right)\frac{\left(k_{B}T\right)^{2}}{n_{1}^{3/2}n_{2}^{3/2}d^{4}}.
\label{eq:large_d}
\end{equation}
It is also usual to express the drag resistivity in this limit in
terms of the Fermi energy, momentum and Thomas-Fermi screening momentum.
To do this we assume a power law energy dispersion relation,
$\e_{\vec{k},c}^{a}(k)=C_{a}k^{\beta_{a}}$, obtaining
\begin{equation}
\rho_{D}=-\frac{\hbar}{e_{1}e_{2}}\frac{\zeta(3)\pi^{2}}{\b_{1}g_{1}\b_{2}g_{2}}\frac{\left(k_{B}T\right)^{2}}{\e_{F1}\e_{F2}}\frac{1}{\left(k_{F1}d\right)\left(k_{F2}d\right)\left(q_{TF1}d\right)\left(q_{TF2}d\right)}.\label{eq:Resistivity}
\end{equation}
For drag between two 2DEG, $\b_{1(2)}=2$, $g_{1(2)}=2$ (spin degeneracy),
we re-obtain the known formula from \cite{Smith1993}. For graphene we obtain exactly the same result, since $\b_{1(2)}=1$, $g_{1(2)}=4$ (spin and valley degeneracy). Finally, for the case where each of
the two layers are formed by graphene bilayers, $\b_{1(2)}=2$, $g_{1(2)}=4$
(spin and valley degeneracy), we have an extra factor of $\frac{1}{4}$.

\subsection{The case of graphene at small interlayer distance}

\begin{figure}
\begin{centering}
\includegraphics[width=10cm]{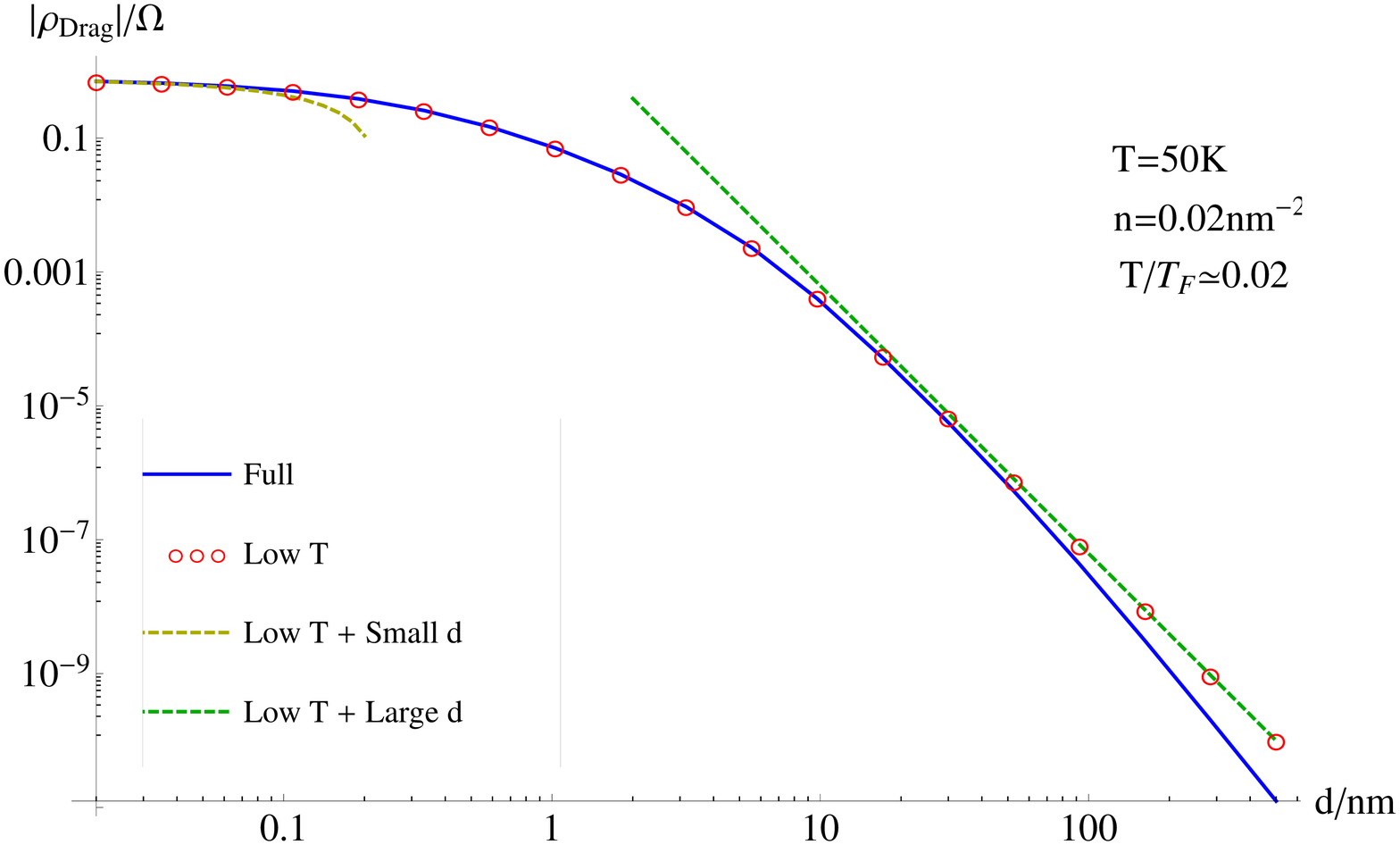}
\par\end{centering}

\caption{\label{fig:Distance}Comparison between computed drag resistivity for
 graphene in vacuum using the full expression from equations \eref{eq:adimnesional}
 and \eref{eq:adimensional int}, and several asymptotic limits as a function of 
the interlayer spacing. The curve \textit{Low T} was computed setting $y=0$ in 
$\e_{RPA}$ and expanding to first order in $y$ the functions $\Phi$ in 
\eref{eq:adimensional int}. The curve \textit{Low T + Small d } was computed 
using \eref{eq:small_d} and \textit{Low T + Large d} using \eref{eq:large_d}. 
We can see that the agreement between the full expression and the low temperature 
one is good for most of the range, but that the small discrepancy is enhanced as $d$ 
increases.} 
\end{figure}

Now we specialize to the case where both metallic layers are formed
by single layer graphene, SLG, and analyze the behaviour of the drag
conductivity when the layer separation is small, $k_{F}d\ll1$. In
this situation we can no longer expand the non-linear susceptibilities
for small $q$ and need to consider its full dependence on $q$. In
graphene the wave function overlap factor is
\begin{equation}
f_{\vec{k},\vec{k}+\vec{q}}^{\l,\l^{\prime}}=\frac{1}{2}
\left(1+\l\l^{\prime}\frac{\vec{k}\cdot\left(\vec{k}+\vec{q}\right)}{\left|\vec{k}\right|\left|\vec{k}+\vec{q}\right|}\right),
\end{equation}
with $\l=+,-$ for the conduction and valence band, respectively.
Therefore, the non-linear susceptibility in the low temperature limit
\eref{eq:NLP_small_omega} reads
\begin{equation}
\vec{\G}_{SLG}(\vec{q},\o)=4\frac{\o\tau_{F}}{\pi\hbar v_{F}}\frac{\vec{q}}{q}\sqrt{1-\left(\frac{q}{2k_{F}}\right)^{2}},\label{eq:NLP_small_omega_SLG}
\end{equation}
where the factor of $4$ comes from the spin and valley degeneracies. \Eref{eq:NLP_small_omega_SLG} is in disagreement with the expressions
obtained in \cite{Hwang2011} both for the momentum independent
and for the linearly momentum dependent transport time cases. However,
we emphasize that in the low temperature limit \eref{eq:NLP_small_omega_SLG}
holds for an arbitrary transport time. Now we notice that for $q<2k_{F}$,
the static polarizability for graphene is given by \cite{Wunsch2006}
\begin{equation}
\chi_{SLG}(q<2k_{F},0)=-\frac{2k_{F}}{\pi\hbar v_{F}}=-q_{TF}\frac{2\e_{0}}{e^{2}}.
\end{equation}
Therefore the dielectric function $\e_{RPA}(q,0)$  still has the form
given by \eref{eq:Thomas-Fermi RPA}, even if we do not assume that $ q $ is small. Since we have $k_{F1(2)}d\ll1$,
we expand to first order in $d$
\begin{eqnarray}
\fl \left|U_{12}(q,0)\right|^{2} & =\left|\frac{V_{12}(q)}{\epsilon_{RPA}(q,0)}\right|^{2}\\
& \simeq\left(\frac{e^{2}}{2\epsilon_{0}}\right)^{2}\left[\frac{1}{(q+q_{TF1}+q_{TF2})^{2}}-d\frac{2(2q_{TF1}q_{TF2}+q(q+q_{TF1}+q_{TF2}))}{(q+q_{TF1}+q_{TF2})^{3}}\right]\nonumber,
\end{eqnarray}
and the drag conductivity becomes
\begin{equation}
\sigma_{D}=\frac{e^{2}}{\hbar}\frac{2^{3}}{3}\a_{g}^{2}
\frac{\tau_{F}^{2}\left(k_{B}T\right)^{2}}{\hbar^{2}}\left[\mathcal{I}^{(0)}\left(k_{F1},k_{F2}\right)-d\mathcal{I}^{(1)}\left(k_{F1},k_{F2}\right)\right],
\end{equation}
where $ \a_{g}=e^{2}/(4\pi\e_{0}v_{F}\hbar) $ is the fine structure constant of graphene and we have defined the functions
\begin{eqnarray}
\fl \mathcal{I}^{(0)}\left(k_{F1},k_{F2}\right)=\int_{0}^{K}\frac{dqq}{\left(q+\left(q_{TF1}+q_{TF2}\right)\right)^{2}}\sqrt{1-\frac{q^{2}}{4k^{2}_{F1}}}\sqrt{1-\frac{q^{2}}{4k^{2}_{F2}}}\\
\fl \mathcal{I}^{(1)}\left(k_{F1},k_{F2}\right)=\int_{0}^{K}dq\frac{2(2q_{TF1}q_{TF2}+q(q+q_{TF1}+q_{TF2}))}{(q+q_{TF1}+q_{TF1})^{3}}\sqrt{1-\frac{q^{2}}{4k^{2}_{F1}}}\sqrt{1-\frac{q^{2}}{4k^{2}_{F2}}}\nonumber,
\end{eqnarray}
where $ K=2\min\left(k_{F1},k_{F2}\right) $, and for graphene $ q_{TF}=4\a_{g}k_{F} $. The drag resistivity becomes
\begin{equation}
\r_{D}=-\frac{\hbar}{e^{2}}\frac{2^{3}\pi^{2}}{3}\frac{\left(k_{B}T\right)^{2}}{\e_{F1}\e_{F2}}\a_{g}^{2}\left[\mathcal{I}^{(0)}\left(k_{F1},k_{F2}\right)-d\mathcal{I}^{(1)}\left(k_{F1},k_{F2}\right)\right].
\label{eq:small_d}
\end{equation}
Identical result for $d=0$ has recently been derived in an independent work \cite{Polini2012}.
For the case where both layers are at the same carrier density $k_{F1}=k_{F2}=k_{F}$ the functions $ \mathcal{I}^{(0)} $ and $\mathcal{I}^{(1)}$ simplify considerably and we obtain,
\begin{eqnarray}
\fl \mathcal{I}^{(0)}\left(k_{F},k_{F}\right)=12\a_{g}-\frac{3}{2}+\left(1-48\a_{g}^{2}\right)\log\left(1+\frac{1}{4\a_{g}}\right)\\
\fl \mathcal{I}^{(1)}\left(k_{F},k_{F}\right)/k_{F}=\frac{2}{3}+44(1-8\a_{g})\a_{g}+\frac{2}{1+4\a_{g}}+32\a_{g}\left(44\a_{g}^{2}-1\right)\log\left(1+\frac{1}{4\a_{g}}\right)\nonumber,
\end{eqnarray}
Therefore, for drag between two SLG layers as the interlayer separation
is increased, the behaviour of the drag resistivity changes 
from a linear dependence on $d$ for $q_{TF}d\ll1$ to a $d^{-4}$ 
dependence at large $q_{TF}d\gg1$. In \fref{fig:Distance}, we can see 
that the small separation expression \eref{eq:small_d} is only 
reliable for $q_{TF}d\lesssim 0.2$ and the large separation expression
 \eref{eq:large_d} for $q_{TF}d\gtrsim 20$. If the graphene layers are 
immersed in a homogeneous dielectric with constant $\e_{r}$ this would 
correspond to $k_{F}d\lesssim 0.02 \e_{r}$ and $k_{F}d\gtrsim 2\e_{r}$, 
respectively. We therefore see, that these limits are not easy to be achieved experimentally.

\section{Drag in graphene: general formula and asymptotic limit
\label{sec:Drag-in-graphene}}

As we have just seen, the dependence on momentum of the transport
time is irrelevant in the low temperature and high density limits, but 
it will be important in general however. The dependence of the transport time
on momentum in graphene depends on the  dominant scattering mechanism.
Both for strong short-range  impurities (resonant scatterers) and Coulomb impurities the
transport time depends linearly on the momentum, $\tau_{\vec{k},\l}=\tau^{0}\left|k\right|$,
where $\tau^{0}$ is a constant with units of length $\times$ time
\cite{Peres2010Review}. The linear dependence of the transport time
is assumed in \cite{Peres2011} and we make the same assumption here.
In this case, the non-linear susceptibility for graphene
becomes:
\begin{eqnarray}
\fl \vec{\G}(\vec{q},\o)=-8\pi v_{F}\tau^{0}\sum_{\l,\l^{\prime}}\int\frac{d^{2}k}{\left(2\pi\right)^{2}}f_{\vec{k},\vec{k}+\vec{q}}^{\l,\l^{\prime}}\left(n_{F}(\e_{\vec{k},\l})-n_{F}(\e_{\vec{k}+\vec{q},\l^{\prime}})\right)\nonumber\\
\times \left(\left(\lambda-\l^{\prime}\right)\vec{k}-\lambda^{\prime}\vec{q}\right)\d\left(\e_{\vec{k},\l}-\e_{\vec{k}+\vec{q},\l^{\prime}}+\hbar\o\right).\label{eq:GrapheneNLP}
\end{eqnarray}
We will follow the steps of \cite{Peres2011} and assume that both
layers are with high electron doping, so that the existence of the
valence band can be ignored. In this case, we  take only the $\l,\l^{\prime}=+,+$
contribution into account. Taking the non-linear susceptibilities at zero temperature, the drag resistivity can be written as

\begin{equation}
\rho_{D}=-\frac{1}{2^{5}}\frac{\hbar}{e^{2}}\frac{\sqrt{\e_{F1}\e_{F2}}}{k_{B}T}\a_{g}^{2}\mathcal{F}\left(k_{F1},k_{F2},d\right).\label{eq:adimnesional}
\label{eq_rhoD_exact}
\end{equation}
The function $\mathcal{F}$ is defined as
\begin{eqnarray}
\fl \mathcal{F}\left(k_{F1},k_{F2},d\right)=\int_{0}^{\infty}dxx^{3}\int_{0}^{\infty}\frac{dy}{\sinh^{2}\left(y\frac{\hbar v_{F}\sqrt{k_{F1}k_{F2}}}{2k_{B}T}\right)}\frac{e^{-2d\sqrt{k_{F1}k_{F2}}x}}{\e_{12}^{2}\abs{\e_{RPA}(x,y)}^{2}}\nonumber\\
\times \frac{\Phi_{1}(x,y)\Phi_{2}(x,y)}{1-\left(\frac{y}{x}\right)^{2}},\label{eq:adimensional int}
\end{eqnarray}
where the functions $\Phi_{a}(x,y)$ are introduced in \ref{sec:Appendix_details} and $x=q/\sqrt{k_{F1}k_{F2}}$ and $y=q/(v_{F}\sqrt{k_{F1}k_{F2}})$. This is exactly the same expression derived in \cite{Peres2011} if one notices that the function $\e(x,y)$ used there is related
to $\e_{RPA}(x,y)$ by $\e(x,y)=x^{2}\left(x^{2}-y^{2}\right)\e_{RPA}(x,y)$.
So far no approximation has been made in the sense that no asymptotic limit has
been considered. A comparison of the different asymptotic behaviours computed
in the previous section with the exact result, \eref{eq_rhoD_exact},
is given in \fref{fig:Distance}.

By the general arguments given in the previous section, the drag
resistivity in the low temperature, high density, large separation
limit should behave as $d^{-4}$. However, in \cite{Peres2011} and 
in this same limit  a dependence of $d^{-6}$ was obtained. In \cite{Hwang2011},
the $d^{-6}$ result is attributed to scaling of the vertex function
$\vec{v}_{\vec{k},\l}\tau_{\vec{k},\tau}$ used in \cite{Peres2011}
with $q^{2}$, while it should scale with $q$ for a constant group
velocity and a transport time linearly dependent on the momentum,
implying that the carrier group velocity used in \cite{Peres2011}
depends linearly on momentum. We clarify that in \cite{Peres2011},
as in this paper, the group velocity used is constant and the transport
time depends linearly in momentum, so that the vertex function depends
linearly in momentum, $\vec{v}_{\vec{k},\l}\tau_{\vec{k},\l}=\lambda v_{F}\tau^{0}\vec{k}$.
However, as we have argued in the previous section, the momentum
dependence of the vertex function is irrelevant in this limit. The
incorrect $d^{-6}$ dependence was obtained in \cite{Peres2011} due
to an error at the end of the asymptotic calculation: 
for small $x$ and $y$ with $y<x$ the function $\Phi(x,y)$
was taken to behave as $\Phi(x,y)\sim\frac{y}{x}$, when it actually
behaves as $\Phi(x,y)\sim\frac{y}{x^{2}}$. This changes the integration
kernel obtained in \cite{Peres2011} in the asymptotic limit
from $q^{5}\sinh^{-2}\left(qd\right)$
to $q^{3}\sinh^{-2}\left(qd\right)$, which changes the dependence
from $d^{-6}$ to the correct behaviour of $d^{-4}$. Therefore, for
$\e_{F1(2)}\beta,\, k_{F1(2)}d\gg1$, the exact \eref{eq:adimnesional}
and \eref{eq:adimensional int}  give 
\begin{equation}
\rho_{D}=-\frac{\hbar}{e^{2}}\frac{\zeta(3)\pi^{2}}{2^{4}}\frac{\left(k_{B}T\right)^{2}}{\e_{F1}\e_{F2}}\frac{1}{\left(k_{F1}d\right)\left(k_{F2}d\right)\left(q_{TF1}d\right)\left(q_{TF2}d\right)},\label{eq:DragSLGlimit}
\end{equation}
in agreement with the general result discussed in the previous section. Expressing \eref{eq:DragSLGlimit}
in terms of the carrier density we get
\begin{equation}
\r_{D}=-\frac{\hbar}{e^{2}}\frac{\zeta(3)}{2^{8}\pi}\frac{1}{\a_{g}^{2}}\frac{\left(k_{B}T\right)^{2}}{\left(v_{F}\hbar\right)^{2}n_{1}^{3/2}n_{2}^{3/2}d^{4}}\,,
\end{equation}
as obtained before, but here we have started from the general expression for the drag, that is,
\eref{eq:adimnesional}. Note that, in this limit the drag 
resistivity decreases as $\alpha_{g}$ is increased. We
can understand this as follows:
 as  $\alpha_{g}$ increases the screening becomes more effective making the 
momentum transfer between layers less effective.

\section{Comparison with experiments\label{sec:Experiments}}

\begin{figure}
\begin{centering}
\includegraphics[width=10cm]{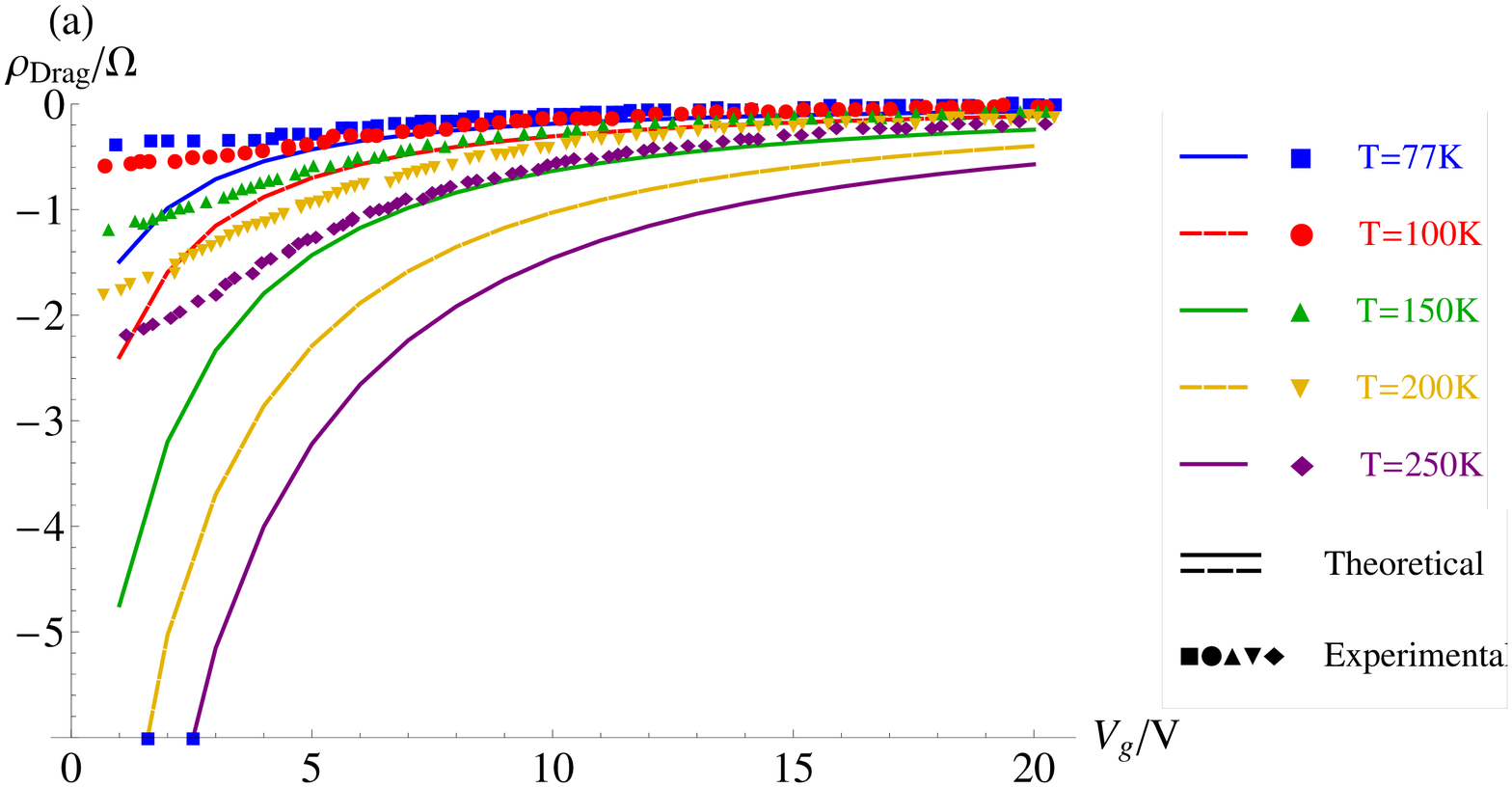}
\par\end{centering}

\begin{centering}
\includegraphics[width=10cm]{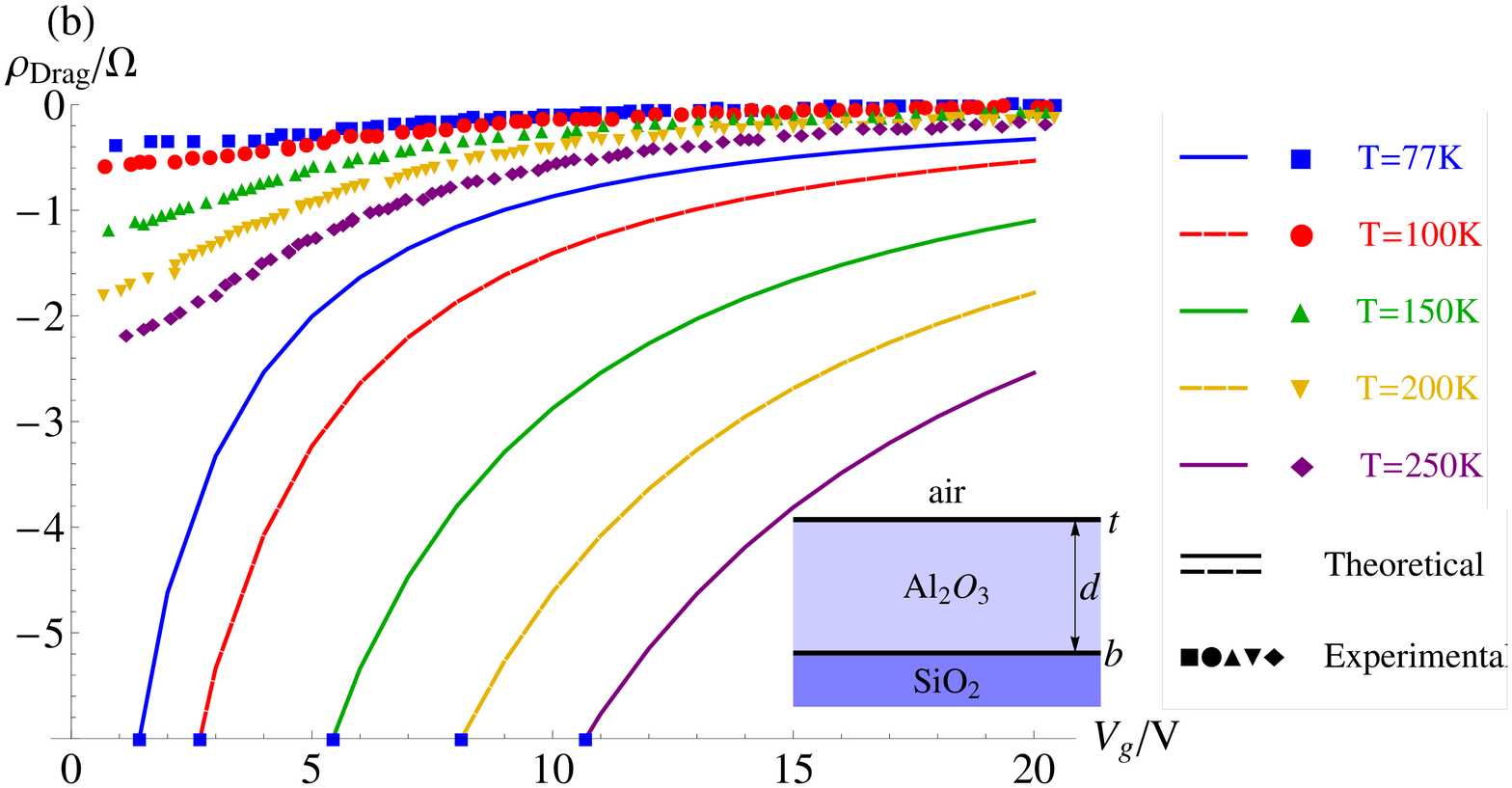}
\par\end{centering}

\begin{centering}
\includegraphics[width=10cm]{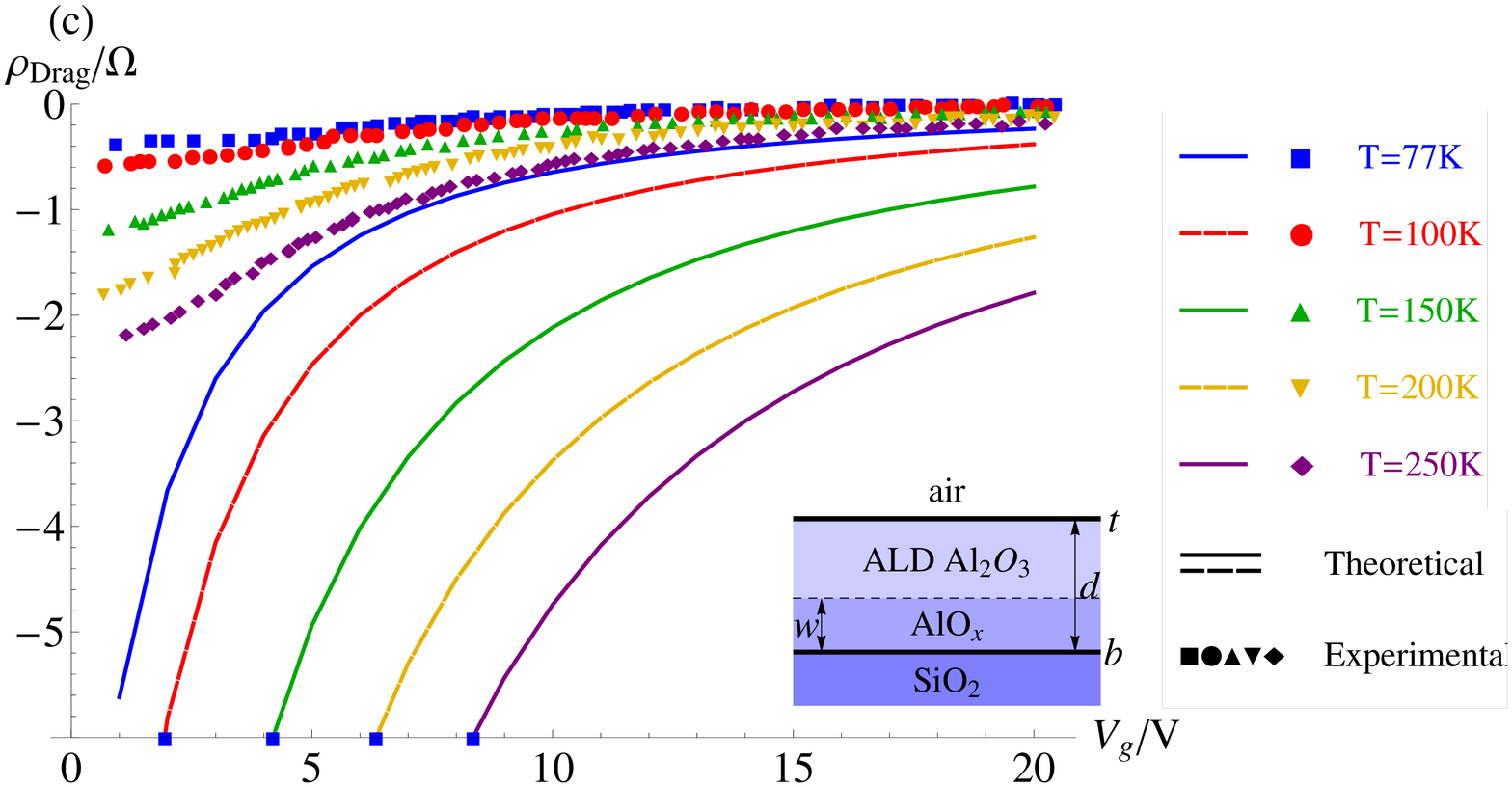}
\par\end{centering}

\caption{\label{fig:DragPlot}Comparison between computed and measured drag
resistivity. The dotted marks are the experimental results from
\cite{Kim2011}. The solid lines are the computed curves. Curves with
the same colour correspond to the same temperature. (a) The computed
curves were calculated using the naive interactions $\left( \e^{(naive)}_{ab}\right) $. 
(b) The computed curves were calculated using the
Coulomb interactions obtained by solving Poisson's equation in a 3 layered dielectric 
depicted in the inset $\left(\e^{(3)}_{ab}(q)\right)$. (c) The computed curves were calculated using the
Coulomb interactions obtained by solving Poisson's equation in a 4 layered dielectric as depicted in the inset $\left(\e^{(4)}_{ab}(q)\right)$.
The value used for the fine structure of graphene, $\a_{g}$, was $\a_{g}=2$ in all
plots. No attempt to fit the data was made.}

\end{figure}

In the experimental setup of \cite{Kim2011} we have two graphene
layers, which we  denote by $t$ (top) and $b$ (bottom). 
Between the two graphene sheets we have a layer
of Al$_{2}$O$_{3}$, thickness $d=d_{t}=7$ nm (in \cite{Peres2011}
it was used a value of $14$ nm). The bottom layer is on 
top of a $d_{b}=280$ nm thickness SiO$_{2}$. Finally,
these layers are on top of a silicon wafer. The carrier density in the
graphene sheets is controlled using a back gate voltage between the
silicon wafer and the bottom layer. For the relation between the
gate voltage and the carrier densities the reader is referred to \cite{Kim2011,Peres2011}.
Given the carrier densities, one can compute the drag resistivity
using \eref{eq:adimnesional} and \eref{eq:adimensional int}. 
For the fine structure constant of graphene we will use 
the accepted value of $\alpha_{g}=2$.  To compute the drag, 
we  must determine the form of the bare Coulomb interactions, $V_{ab}(\vec{q})$ \eref{eq:bareCoulomb}, with $a,b=t,b$. 
In a naive treatment of the Coulomb interactions one could write \cite{Peres2011}:
\begin{eqnarray}
\e^{(naive)}_{tt}=\frac{\e_{air}+\e_{Al_{2}O_{3}}}{2}
~~~~~\e^{(naive)}_{tb}=\e_{Al_{2}O_{3}}~~~~~\e^{(naive)}_{bb}
=\frac{\e_{Al_{2}O_{3}}+\e_{SiO_{2}}}{2} 
\label{eq:Naive}
\end{eqnarray}
with $\e_{air}$, $\e_{Al_{2}O_{3}}$ and $\e_{SiO_{2}}$ the 
dielectric constants of air, Al$_{2}$O$_{3}$ and SiO$_{2}$, 
respectively. Using the values of $\e_{air}=1$, $\e_{SiO_{2}}=3.8$ 
and $\e_{Al_{2}O_{3}}=5.6$ \cite{Tutuc2012dieletric} we computed the
 drag resistivity curves with this model for the interactions. 
In \fref{fig:DragPlot}(a), we can see the comparison between 
the experimental results from \cite{Kim2011} and this calculation.
 Clearly the agreement is not very good. Although \eref{eq:Naive}
 is correct is some limits, to obtain the exact form of the Coulomb 
interactions in layered dielectric medium one must solve Poisson's 
equation (see \ref{sec:Coulomb_Interaction}). For a 3 layered dielectric 
structure, for the effective dielectric constants $\e_{ab}(q)$ in
 \eref{eq:bareCoulomb} we would have to use the functions 
$\e^{(3)}_{ab}(q)$ defined in \ref{sec:Coulomb_Interaction}, 
with $\e_{1}=\e_{air}=1$, $\e_{2}=\e_{Al_{2}O_{3}}=5.6$ and
 $\e_{3}=\e_{SiO_{2}}=3.8$. In \fref{fig:DragPlot}(b), 
we can see the comparison between the computed resistivity 
using this interactions and the experimental results. Although 
this is a more rigorous treatment for the interactions than the 
naive one, \eref{eq:Naive}, the computed drag curves deviate 
even further from the experimental results. One could proceed  as in \cite{Peres2011} and use $\a_{g}$ as a fitting parameter. We found, 
however, that it would be necessary a value of the order $\a_{g}\sim5$ 
and $\a_{g}\sim16$,  for the interactions with $\e^{(naive)}_{ab}$ and 
$\e^{(3)}_{ab}$, respectively, to fit the data; both values are unrealistic. 
Taking into account that the SiO$_{2}$ layer is finite when solving Poisson's 
equation  has virtually no effect in the computed drag resistivity. This situation 
is puzzling at least. However, more attention must be paid to how the devices 
from \cite{Kim2011} are constructed. The Al$_{2}$O$_{3}$ dielectric layer is 
deposited in two steps. (i) First a $2$ nm aluminium layer is deposited on 
top of the bottom graphene layer, being later oxidized. We will refer to 
this as AlO$_{x}$ layer. (ii) On top of the oxidized aluminium layer, 
a $5$ nm Al$_{2}$O$_{3}$ layer is directly deposited through atomic 
layer deposition. We will refer to this as ALD layer. Completed the 
procedure a $7$ nm Al$_{2}$O$_{3}$ is obtained. The reason the Al$_{2}$O$_{3}$ 
is not directly deposited on top of graphene is due to graphene being chemically inert. 
In \cite{Tutuc2012dieletric}, the dielectric constant of the ALD alumina was 
determined by studying how the capacitance of devices similar to the ones 
from \cite{Kim2011} scales with the tickness of the ALD layer keeping the 
AlO$_{x}$ layer thickness fixed, obtaining the already referred value of 
 $\e_{Al_{2}O_{3}}=5.6$. The AlO$_{x}$ layer contributed with a finite 
capacitance, which corresponds to a dielectric constant of $\e_{AlO_{x}}=2.7$. 
The reason why $\e_{AlO_{x}}$ differs from $\e_{Al_{2}O_{3}}$ is not clear
 but it is most likely due to interface effects between the graphene layer and
 the dielectric or between the ALD and the AlO$_{x}$ layers. In this case, it is 
not clear if we can attribute a bulk value for $\e_{AlO_{x}}$. Nevertheless, we 
still considered this situation. For this situation of a 4 layered dielectric we
 use the functions $\e^{(4)}_{ab}(q)$ from \ref{sec:Coulomb_Interaction}, now 
with $\e^{\prime}_{2}=\e_{AlO_{x}}=2.7$ and $w=2$ nm. The comparison between 
 the drag curves computed this way and the experimental data can be seen in 
\fref{fig:DragPlot}(c). We can see that there is a better approximation 
with the experimental results, but this should be regarded with caution.

\section{Conclusions}
In this paper we have showed that in the limit of low temperature, high density, large interlayer distance and strong screening, the drag resistivity should always behave with $T^{2}$ and $d^{-4}$. This result was obtained for general dispersion relation, momentum dependence of the transport time and electronic wave function structure. It is therefore a more general result than previous ones. Central to this fact is the general expression derived for the non-linear susceptibility, \eref{eq:NLP_small_omega}, in the low temperature limit to lowest order in frequency. Being general, this result also applies to graphene, and should close the ongoing debate regarding the behaviour of drag for this system in the aforementioned limit. We also derived an asymptotic expression for drag in graphene in the limit of  low temperature, high density and small layer separation. Finally, we compared the available experimental data on drag in graphene from \cite{Kim2011} with our theoretical model and found out that if one insists in the conventional value for graphene fine structure constant, $\alpha_{g}=2$, a quantitative understanding of the data is still lacking. This is most likely due to the complexity of the dielectric substrate used and one hopes that the theoretical model will be quantitatively more successful in devices built using simpler substrates, such as boron nitride \cite{Ponomarenko2011}. Another possibility to account for this discrepancy could be spacial inhomogeneities in the chemical potential of the graphene layers, which are not taken into account with the present formalism. Although, it is not clear if these should be important in the experimental range investigated, spacial inhomogeneities in graphene's carrier density will for sure play an important role in the electron-hole puddle 
regime and should be object of future work.

\ack

We thank useful discussions with F. Guinea, J.M.B. Lopes dos Santos and
A.H. Castro Neto. Special thanks to E. Tutuc for helpful discussions regarding the construction of graphene double layer structures. B. Amorim was supported by Funda\c{c}\~{a}o
para a Ci\^{e}ncia e a Tecnologia (FCT) through Grant No. SFRH/BD/78987/2011.

\appendix

\section{Electron-electron Coulomb interaction}\label{sec:Coulomb_Interaction}

To obtain the correct form of the electron-electron interaction in a dielectric material one must solve Poisson's equation, $-\nabla(\epsilon \nabla \phi)=\rho_{free}/\epsilon_{0}$, where $\rho_{free}$ is the free charge density. We are interested in situations where the dielectric constant is a piecewise constant function of the $z$ coordinate, with discontinuities at positions $z_{i}$. Since, we want the interaction matrix element between states with well defined momentum in the $x,y$ directions and well defined position in the $z$ direction, it is useful to introduce $\phi(q,z)=\int d^{2}x\phi(\vec{x},z)e^{-i \vec{q}\cdot\vec{x}}$, where $\vec{q}=(q_{x},q_{y})$ and $\vec{x}=(x,y)$. For a point charge located at $(\vec{x},z)=(\vec{0},z_{p})$, $\rho_{free}=e\delta(\vec{0})\delta(z_{p})$, we obtain the following equation
\begin{equation}
-\frac{\pd}{\pd z}\left(\epsilon(z)\frac{\pd}{\pd z}\phi(q,z)\right)+q^{2}\phi(q,z)=\frac{e}{\epsilon_{0}} \delta(z_{p}).
\label{eq:Poisson_eq}
\end{equation}
The potential $\phi(q,z)$ is continuous everywhere, and the the function $\epsilon(z)\frac{\pd\phi(q,z)}{\pd z}$ is continuous except at the position of the point charge. Therefore we have the boundary conditions
\begin{eqnarray}
\phi(q,z_{i}^{+}) = \phi(q,z_{i}^{-}),\nonumber\\
\epsilon(z_{i}^{-})\frac{\pd\phi(q,z_{i}^{-})}{\pd z} - \epsilon(z_{i}^{+})\frac{\pd\phi(q,z_{i}^{+})}{\pd z} = \frac{e}{\epsilon_{0}}\delta_{z_{i},z_{p}}\,.
\label{eq:boundary} 
\end{eqnarray}
Solving equation \eref{eq:Poisson_eq} together with the boundary conditions \eref{eq:boundary} and imposing that $\phi(q,z)$ decays at infinity, we can obtain the potential created by the point charge and from that the bare electron-electron Coulomb interaction. Let us consider that the metallic plates are located at $z=0$ and $z=d$. The electron-electron interaction can be cast in the form given in \eref{eq:bareCoulomb} with $a,b=t,b$, where $t$ refers to the top layer and $b$ refers to the bottom layer. For a 3 layered dielectric:
\begin{equation}
\e^{(3)}(z)=\cases{
					\e_{1}&, z $>$ d\\
					\e_{2}&, d $>$ z $>$ 0,\\
					\e_{3}&, 0 $>$ z}
\end{equation}
solving Poisson's equation gives
\begin{eqnarray}
\e^{(3)}_{tt}(q) & = & \frac{e^{2qd}(\e_{3}+\e_{2})(\e_{1}+\e_{2})-(\e_{3}-\e_{2})(\e_{1}-\e_{2})}{2\left(1+e^{2qd}\right)\e_{2}-2\left(1-e^{2qd}\right)\e_{3}},\nonumber \\
\e^{(3)}_{tb}(q) & = & \frac{e^{2qd}(\e_{1}+\e_{2})(\e_{2}+\e_{3})-(\e_{1}-\e_{2})(\e_{3}-\e_{2})}{4e^{2qd}\e_{2}},\\
\e^{(3)}_{bb}(q) & = & \frac{e^{2qd}(\e_{1}+\e_{2})(\e_{3}+\e_{2})-(\e_{1}-\e_{2})(\e_{3}-\e_{2})}{2\left(1+e^{2qd}\right)\e_{2}-2\left(1-e^{2qd}\right)\e_{1}}.\nonumber 
\end{eqnarray}
This result was previously given in \cite{Katsnelson2011} and \cite{Profumo2010}. If we consider a 4 layered dielectric,
\begin{equation}
\e^{(4)}(z)=\cases{
					\e_{1}&, z $>$ d\\
					\e_{2}&, d $>$ z $>$ w\\
					\e^{\prime}_{2}&, w $>$ z $>$ 0\\
					\e_{3}&, 0 $>$ z},
\end{equation}
we obtain,
\begin{eqnarray}
\fl \e^{(4)}_{tt}(q) = \frac{D^{(4)}(q)}{2}\left[
e^{2q(d+w)}\left(\epsilon_{2}+\epsilon_{2}^{\prime}\right)\left(\epsilon_{3}+\epsilon_{2}^{\prime}\right)
+e^{2dq}\left(\epsilon_{2}-\epsilon_{2}^{\prime}\right)\left(\epsilon_{2}^{\prime}-\e_{3}\right)\right.\nonumber\\
\left. +e^{4qw}\left(\epsilon_{2}-\epsilon_{2}^{\prime}\right)\left(\epsilon_{3}+\epsilon_{2}^{\prime}\right)
+e^{2qw}\left(\epsilon_{2}+\epsilon_{2}^{\prime}\right)\left(\epsilon_{2}^{\prime}-\e_{3}\right)\right]^{-1},\nonumber \\
\fl \e^{(4)}_{tb}(q) = \frac{D^{(4)}(q)}{8e^{2q(d+w)}\epsilon_{2}\e_{2}^{\prime}},\\
\fl \e^{(4)}_{bb}(q) = \frac{D^{(4)}(q)}{2}\left[
e^{2q(d+w)}\left(\epsilon_{1}+\epsilon_{2}\right)\left(\epsilon_{2}+\epsilon_{2}^{\prime}\right)
-e^{2dq}\left(\epsilon_{1}+\epsilon_{2}\right)\left(\epsilon_{2}-\epsilon_{2}^{\prime}\right)\right.\nonumber\\
\left. +e^{4qw}\left(\epsilon_{2}-\epsilon_{2}^{\prime}\right)\left(\epsilon_{3}+\epsilon_{2}^{\prime}\right)
-e^{2dq}\left(\epsilon_{1}+\epsilon_{2}\right)\left(\epsilon_{2}-\epsilon_{2}^{\prime}\right)\right]^{-1},\nonumber 
\end{eqnarray}
where $D^{(4)}(q)$ is defined as
\begin{eqnarray}
\fl D^{(4)}(q) =
e^{2q(d+w)}\left(\epsilon_{1}+\epsilon_{2}\right)\left(\epsilon_{2}+\epsilon_{2}^{\prime}\right)\left(\epsilon_{2}^{\prime}+\epsilon_{3}\right) +e^{2dq}\left(\epsilon_{1}+\epsilon_{2}\right)\left(\epsilon_{2}-\epsilon_{2}^{\prime}\right)\left(\epsilon_{2}^{\prime}-\epsilon_{3}\right)\nonumber\\
+e^{4qw}\left(\epsilon_{1}-\epsilon_{2}\right)\left(\epsilon_{2}-\epsilon_{2}^{\prime}\right)\left(\epsilon_{2}^{\prime}+\epsilon_{3}\right)
+e^{2qw}\left(\epsilon_{1}-\epsilon_{2}\right)\left(\epsilon_{2}+\epsilon_{2}^{\prime}\right)\left(\epsilon_{2}^{\prime}-\epsilon_{3}\right).
\end{eqnarray}

\section{Details of the computation of the drag resistivity in graphene\label{sec:Appendix_details}}

The contribution to the non-linear susceptibility of graphene \eref{eq:GrapheneNLP}
coming only from the conductance band, $\l,\l^{\prime}=+,+$ is

\begin{eqnarray}
\fl \vec{\G}(\vec{q},\o)  & =  8\pi v_{F}\tau^{0}\vec{q}\int\frac{d^{2}k}{\left(2\pi\right)^{2}}f_{\vec{k},\vec{k}+\vec{q}}^{+,+}& \left(n_{F}(\e_{\vec{k},+})-n_{F}(\e_{\vec{k}+\vec{q},+})\right)
\d\left(\e_{\vec{k},+}-\e_{\vec{k}+\vec{q},+}+\hbar\o\right)\nonumber\\
\fl  & =  -2v_{F}\tau^{0}\vec{q}\textrm{Im}\chi^{++}\left(q,\omega\right)
\end{eqnarray}
where $\chi^{++}(q,\o)$ is the contribution to the graphene polarizability
coming only from the conductance band,
\begin{equation}
\chi^{++}(q,\o)=4\int\frac{d^{2}k}{\left(2\pi\right)^{2}}f_{\vec{k},\vec{k}+\vec{q}}^{+,+}\frac{n_{F}(\e_{\vec{k},+})-n_{F}(\e_{\vec{k}+\vec{q},+})}{\e_{\vec{k},+}-\e_{\vec{k}+\vec{q},+}+\hbar\o+i0^{+}}.
\end{equation}
At zero temperature, this can be computed analytically and the result can be written as
\begin{equation}
\textrm{Im}\chi^{++}(\vec{q},\o)=\frac{1}{4\pi v_{F}\hbar}\frac{q}{\sqrt{1-\left(\frac{\o}{qv_{F}}\right)^{2}}}\Phi(q,\o),
\end{equation}
where the function $\Phi(q,\o)$ is defined as
\begin{eqnarray}
\fl \Phi(q,\omega>0)=  \Phi^{+}(q,\o)\Theta\left(\frac{\o}{v_{F}}-q+2k_{F}\right)\Theta\left(q-\frac{\o}{v_{F}}\right)\nonumber\\
+\Phi^{-}(q,\o)\Theta\left(k_{F}-\frac{\o}{v_{F}}-\left|k_{F}-q\right|\right),
\end{eqnarray}
where
\begin{eqnarray}
\fl \Phi^{\pm}(q,\o)  =  \pm \cosh^{-1}\left(\frac{2k_{F} \pm \o/v_{F}}{q}\right)\mp\frac{2k_{F}\pm\o/v_{F}}{q}\sqrt{\left(\frac{2k_{F}\pm\o/v_{F}}{q}\right)^{2}-1}.
\end{eqnarray}
The drag conductivity therefore becomes
\begin{equation}
\sigma_{D}=\frac{e^{2}(\tau^{0})^2}{2^{7}\pi^{4}\hbar^{3}k_{B}T}\int_{0}^{\infty}dqq^{5}\int_{0}^{\infty}\frac{d\o \abs{U_{12}(\vec{q},\o)}^{2}}{\sinh^{2}\left(\beta\hbar\omega/2\right)}\frac{\Phi_{1}(q,\o)\Phi_{2}(q,\o)}{1-\left(\frac{\o}{qv_{F}}\right)^{2}}.
\end{equation}
The intralayer graphene conductivity of layer $a$ at low temperature is given by $\sigma_{aa}=e^{2}v_{F}n_{a}\tau^{0} / \hbar=e^{2}\e_{F}\tau_{F}/(\pi\hbar^{2})$
where $k_{Fa}\tau^{0}=\tau_{Fa}$ is the transport time at the Fermi level. We write the interlayer interaction as $U_{12}=e^{2}\exp(-qd)/(2\epsilon_{RPA}(\vec{q},\omega)\epsilon_{12}(q)\epsilon_{0}q)$. Introducing the adimensional quantities, $x=q/\sqrt{k_{F1}k_{F2}}$ and $y=\omega /(v_{F} \sqrt{k_{F1}k_{K2}})$, the drag resistivity becomes
\begin{eqnarray}
\fl \rho_{D}=-\frac{1}{2^{5}}\frac{\hbar}{e^{2}}\frac{\sqrt{\e_{F1}\e_{F2}}}{k_{B}T}\a_{g}^{2}\int_{0}^{\infty}  dxx^{3}\int_{0}^{\infty}\frac{dy}{\sinh^{2}\left(y\frac{\hbar v_{F}\sqrt{k_{F1}k_{F2}}}{2k_{B}T}\right)} \nonumber\\
\times \frac{e^{-2d\sqrt{k_{F1}k_{F2}}x}}{\e_{12}(q)^{2}\abs{\e_{RPA}(x,y)}^{2}}\frac{\Phi_{1}(x,y)\Phi_{2}(x,y)}{1-\left(\frac{y}{x}\right)^{2}},
\end{eqnarray}
where $\alpha_{g}$ is the fine structure constant of graphene.

\section*{References}


\providecommand{\newblock}{}

\end{document}